\documentclass[letter]{article}
\usepackage{booktabs}   
\usepackage{subcaption} 
\usepackage[utf8]{inputenc}
\usepackage{enumitem}
\usepackage{tikz}
\usetikzlibrary{arrows}
\usetikzlibrary{shadows}
\usetikzlibrary{shapes}
\usetikzlibrary{backgrounds}
\usepackage{verbatim}
\usepackage{fancyvrb}
\usepackage{csquotes}
\usepackage[htt]{hyphenat}
\usepackage{balance}
\usepackage{url}
\usepackage{hyperref}
\usepackage[left=1in,right=1in,top=1in,bottom=1in]{geometry}
\title{Verifying x86 Instruction Implementations}
\author{Shilpi Goel \qquad  Anna Slobodova \qquad Rob Sumners \qquad Sol Swords \\
\small{Centaur Technology, Inc., Austin, TX. USA}\\
\small{{\tt \{shilpi,anna,rsumners,sswords\}@centtech.com}}}
\date{}
\begin{document}

\maketitle

\begin{abstract}
  Verification of modern microprocessors is a complex task that
  requires a substantial allocation of resources. Despite significant
  progress in formal verification, the goal of complete
  verification of an industrial design has not been achieved. In this
  paper, we describe a current contribution of formal methods to the
  validation of modern x86 microprocessors at Centaur Technology.  We
  focus on proving correctness of instruction implementations, which
  includes the decoding of an instruction, its translation into a
  sequence of micro-operations, any subsequent execution of traps to
  microcode ROM, and the implementation of these micro-operations in
  execution units.  All these tasks are performed within one
  verification framework, which includes a theorem prover, a verified
  symbolic simulator, and SAT solvers. We describe the work of
  defining the needed formal models for both the architecture and
  micro-architecture in this framework, as well as tools for
  decomposing the requisite properties into smaller lemmas which can
  be automatically checked. We additionally cover the advantages and
  limitations of our approach. To our knowledge, there are no similar
  results in the verification of implementations of an x86
  microprocessor.
\end{abstract}

\section{Introduction}\label{sec:introduction}

The capacity of formal verification tools and techniques has improved
significantly in recent decades. The capabilities of efficient solvers for
decidable logic problems have extended the umbrella of tasks that can
be discharged automatically with proper setup.
In addition, greater computational capacity with faster machines and more
parallel and distributed processing have given more leverage to formal tools.

Unfortunately, the scale and complexity of microprocessor designs have kept
ahead of the ability to formally verify their correctness.  Modern
microprocessors support more complex and extensive instruction sets, more
threads and cores, and more aggressive optimizations to reduce latency and
increase parallelism. In addition, as microprocessors develop into
systems-on-a-chip, the potential for further optimizations and complexity only
grows. Indeed, the best approach for applying formal verification resources in
microprocessor verification is divide-and-conquer. Splitting an unmanageable
task into smaller ones not only limits the scope in which verification is
carried out, but also allows for tackling different parts of the design with
different methods.  It is a big advantage when all those different methods can
be implemented within the same formalism, ensuring consistent 
composition of intermediate results.

Our work presented in this paper describes attempts to prove correctness of
instruction implementations for an industrial x86 processor.
This task includes verification of the
Register-Transfer Level (RTL) design implementation of instruction decode,
translation, and execution units as well as associated microcode. Verification
of some of these aspects of an x86 instruction execution have been previously
reported\cite{jones2002symbolic, aagaard1999formal, russinoff2019formal,
  intel-i7-fv, memocode11, acl2-industrial-2015}.
However, 
those efforts considered those steps in isolation (one at a time) and
none of them incorporated all of them within one verification
framework.  Compared to our previous
work~\cite{HS09,memocode11,centaur-itp14,sat15,spISA2019}, we have:
\begin{itemize}

\item improved verification of decoding and translation by supporting
the proof of a more general form of an instruction;

\item added verification of microcode generated as a result of
  instruction decoding and increased automation of these proofs;

\item expanded the verification coverage by
considering {\em all} micro-operations dispatched to execution units and
unifying the scope of proofs to the same top-level execution module.
\end{itemize}
The largest remaining gaps in our correctness proofs lie in the verification
of memory access and scheduling of micro-operations.

We use the ACL2 theorem prover \cite{acl2:home} and built-in verified proof
routines \cite{10-swords-dissertation,17-swords-term-level} to model and verify
a target RTL implementation of instruction execution. All specifications are
written in ACL2---our x86 ISA specification
\cite{goel-dissertation,32-bit-x86isa} as well as
our proprietary micro-architectural model.
Leveraging publicly available ACL2 libraries (many of which were developed by
our team), we translate the SystemVerilog design into a formal model within
ACL2.

This paper is organized as follows. Section~\ref{sec:motivation} will
cover preliminary topics of x86 instruction set architecture as well
as micro-architecture. We present an overview of our approach and the
properties we target in Section~\ref{sec:overview}. Our methodology is
explained using examples that describe the verification of two x86
instructions.  In Sections~\ref{sec:shrd} and~\ref{sec:vpshrdq}, we
present detailed example instruction proofs and, in particular, we
cover microcode definitions and verification. We present additional
notes on how we generate proofs, and perform proof exploration and
debugging in Section~\ref{sec:automate}. We discuss related work in
Section~\ref{sec:related} and conclude the paper in
Section~\ref{sec:future}.

\section{Preliminaries}\label{sec:motivation}


The primary goal of formal verification efforts at Centaur Technology is to prove that
microprocessor designs operate in accordance with their specifications. The
primary component of the specification for these microprocessors is compliance
with the x86 instruction set architecture. We discuss instruction set
architectures, x86 in particular, and the common aspects of microprocessor
implementations of complex instruction sets like x86. 


\subsection{x86 Instruction Set Architecture}\label{sec:x86isa}


Microprocessors execute programs, which are defined as a sequence of
instructions. These instructions result from the parsing or decoding of byte
sequences fetched from memory. Each instruction performs operations on data
loaded from either registers or memory, and stores the results in either
registers or memory. An {\em Instruction Set Architecture} (or ISA) specifies
how byte sequences decode into legal instructions and how the resulting
instructions operate on the current programmer-visible state. An {\em ISA
  model} formally defines the semantics of an ISA.  It generally consists of
a type definition for the programmer-visible state and an update function,
which takes the current state and returns the updated state after executing an
instruction. 


The x86\cite{intel-x86-doc}
family of instruction set architectures covers a wide range of microprocessors
built by several companies, including Centaur Technology, over the last several decades. The
x86 family began as a 16-bit architecture and was incrementally extended to
support wider data paths and an ever-increasing library of operations. We use
``the x86 ISA'' to mean
the current iteration of x86, which supports legacy 16-bit and
32-bit modes, as well as modern 64-bit modes and vector instructions supporting
up to 512-bit operations.

Figure~\ref{fig:x86-isa-model} presents the state type and update
function of the x86 ISA model, which we call {\tt x86isa}~\cite{x86isa-github}, that is
used as the specification in our project. The state of the {\tt
  x86isa} consists of an instruction pointer (IP), general-purpose
registers, and a model of memory. In addition, the {\tt x86isa} state
includes a bank of configuration registers controlling modes of
operation, additional registers and memory tables to define address
mappings, larger registers for vector operations, and registers which
store side results and effects from instruction operation. Even
relatively simple x86 integer instructions can be complicated, with
conditions on how to map address accesses to memory and proper
updating of flags denoting edge conditions in instruction
operation.

The {\tt x86isa} update function {\tt x86-model-step} (see
Figure~\ref{fig:x86-isa-model}) is a composition of four component
functions. The function {\tt x86-fetch-code} pulls bytes from memory at the
current instruction pointer. The function {\tt x86-decode} takes the fetched
bytes and current configuration, and returns an exception {\it dx} and the
instruction {\it instr} to execute.  {\it dx} is relevant when a byte
sequence is ill-formed or when the instruction is illegal to execute in the
current state.  The instruction {\it instr} is relevant when there is no
exception.  The function {\tt x86-exec} takes the valid instruction {\it instr}
and returns its computed results {\it rslts} and exceptions {\it ex}, if
any occur during the computation.  The results {\it rslts} include output data
as well as side effects of the computation. For instance, many
instructions will update flags to denote boundary conditions that have been
reached---e.g., a zero flag is hit on a {\em decrement} instruction if the
result is zero; this can be checked as a condition for, say, branching out of a
loop. Finally, the function {\tt x86-update} updates the x86 state either with
the results of the computation or the effects of a triggered exception, and
this state is then returned from the update function.

\begin{figure}[ht]
\begin{tabbing}
\hspace{.07 in} \= $\langle$ex, rslts$\rangle$ \= \hspace{.1 in} \= \kill
{\bf type} {\tt x86-model-state} {\bf is} \\
\> {\bf tuple} $\langle${\it IP, config, registers, memory}$\rangle$ \\
\\
{\bf func} {\tt x86-model-step} {\it (state)} {\bf is} \\
\> {\it bytes} \> = \> {\tt x86-fetch-code}{\it(state.IP, state.memory)} \\
\> $\langle${\it dx, instr}$\rangle$ \> = \> \texttt{\textbf{x86-decode}}{\it(bytes, state.config)} \\
\> $\langle${\it ex, rslts}$\rangle$ \> = \> \texttt{\textbf{x86-exec}}{\it(instr, state)} \\
\> {\it next-state} \> = \> {\tt x86-update}{\it(dx, ex, rslts, state)} \\
\> {\bf return} {\it next-state} 
\end{tabbing}
\caption{x86 ISA Model. Functions in bold represent
specifications which we use for proofs about the microprocessor
design. Details are in Figure~\ref{fig:single-exec}.}
\label{fig:x86-isa-model}
\end{figure}

\begin{figure}[ht]
\centering
\begin{tikzpicture}

  \draw (0,1) rectangle (2,2);
  \node at (1,1.7) {{\tt \small instruction}};
  \node at (1,1.3) {{\tt \small prefixes}};

  \draw (2,1) rectangle (3.5,2);
  \node at (2.75,1.7) {{\tt \small opcode}};
  \node at (2.75,1.3) {{\tt \small bytes}};

  \draw (3.5,1) rectangle (5.5,2);
  \node at (4.5,1.7) {{\tt \small source/}};
  \node at (4.5,1.3) {{\tt \small destination}};

  \draw (5.5,1) rectangle (7.5,2);
  \node at (6.5,1.7) {{\tt \small immediate}};
  \node at (6.5,1.3) {{\tt \small data}};

  \draw [>=stealth,->] (2,.75) -- (2.5,.75);
  \draw [>=stealth,->] (2.5,.75) -- (2,.75);

  \node at (3.75,0.5) {{\tt instr. decoded from 1 byte up to 15 bytes}};

  \draw [>=stealth,->] (0,.25) -- (7.5,.25);
  \draw [>=stealth,->] (7.5,.25) -- (0,.25);

  \path[draw,dashed,dotted] (0,.25) to (0,1);
  \path[draw,dashed,dotted] (7.5,.25) to (7.5,1);

  \path[draw,dashed,dotted] (2,.75) to (2,1);
  \path[draw,dashed,dotted] (2.5,.75) to (2.5,1);
\end{tikzpicture}
\caption{Decode of x86 instr. from byte sequence}
\label{fig:inst-pieces}
\end{figure}
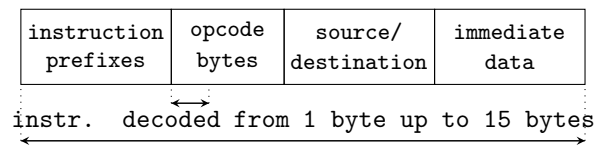

The decoding and execution of even a single x86 instruction is complex.  To
begin with, instructions can be encoded in as few as 1 byte and as many as 15
bytes. Figure~\ref{fig:inst-pieces} presents the pieces of an x86 instruction,
which consists of a variable number of bytes for instruction prefixes (i.e.,
legacy prefix overrides, and REX, VEX, and EVEX prefix bytes), up to 3 bytes
for the instruction opcode, a variable number of bytes that specify which
registers or memory addresses will be sources and destinations of the
instruction (i.e., the ModR/M byte, SIB byte, and displacement bytes), and up
to 8 bytes defining immediate data used in the instruction.

With the extensive set of operations performed by x86 instructions, and the
various modes and configurations in which these operations are performed, a
complete formal x86 ISA model is a significant undertaking in and of
itself~\cite{goel-dissertation}. We derive our specification for x86
instruction decode and execution from existing work~\cite{goel-dissertation,
  centaur-itp14}, but with several necessary modifications and extensions.
We recognize around 3400 x86 instruction {\em variants} in our formal models.  An
instruction with the same mnemonic can have different variants--e.g., one
variant can have a register as the destination operand and another can have a
memory location.


\subsection{Microprocessor Organization}


Microprocessors which implement complex instruction architectures like x86
generally do not directly execute instructions, but instead, translate them
into sequences of simpler operations, termed {\em micro-operations} or {\em
  uops}.  These uops in effect define a new internal instruction architecture
within the microprocessor, termed {\em micro-architecture}. The
micro-architecture is optimized to support pipelining and the scheduling and
mapping of operations onto execution units, which carry out the computations of
these uops.  Note that different variants of the same instruction may
internally correspond to different uops.

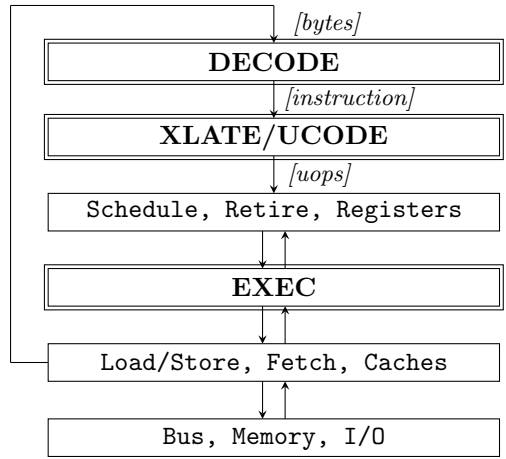
\begin{figure}[ht]
\centering
\begin{tikzpicture}

  \node at (4.7,5.75) {{\small \em [bytes]}};

  \draw (1,5) rectangle (7,5.5);
  \draw (.95,4.95) rectangle (7.05,5.55);
  \node at (4,5.25) {{\tt \bf DECODE}};

  \draw [>=stealth,->] (4,5) -- (4,4.5);
  \node at (5,4.75) {{\small \em [instruction]}};

  \draw (1,4) rectangle (7,4.5);
  \draw (.95,3.95) rectangle (7.05,4.55);
  \node at (4,4.25) {{\tt \bf XLATE/UCODE}};

  \draw [>=stealth,->] (4,4) -- (4,3.5);
  \node at (4.6,3.75) {{\small \em [uops]}};

  \draw (1,3) rectangle (7,3.5);
  \node at (4,3.25) {{\tt Schedule, Retire, Registers}};

  \draw [>=stealth,->] (3.85,3) -- (3.85,2.5);
  \draw [>=stealth,->] (4.15,2.5) -- (4.15,3);

  \draw (1,2) rectangle (7,2.5);
  \draw (.95,1.95) rectangle (7.05,2.55);
  \node at (4,2.25) {{\tt \bf EXEC}};

  \draw [>=stealth,->] (3.85,2) -- (3.85,1.5);
  \draw [>=stealth,->] (4.15,1.5) -- (4.15,2);

  \draw (1,1) rectangle (7,1.5);
  \node at (4,1.25) {{\tt Load/Store, Fetch, Caches}};

  \draw [>=stealth,->] (3.85,1) -- (3.85,.5);
  \draw [>=stealth,->] (4.15,.5) -- (4.15,1);

  \draw (1,0) rectangle (7,0.5);
  \node at (4,0.25) {{\tt Bus, Memory, I/O}};

  \draw [>=stealth,-] (1,1.25) -- (.5,1.25);
  \draw [>=stealth,-] (.5,1.25) -- (.5,6);
  \draw [>=stealth,-] (.5,6) -- (4,6);
  \draw [>=stealth,->] (4,6) -- (4,5.5);

\end{tikzpicture}
\caption{Microprocessor Organization}
\label{fig:microproc-org}%
\end{figure}

Figure~\ref{fig:microproc-org} presents a standard generalized architecture of
an x86 microprocessor. Program code to execute is loaded from memory through a
cache hierarchy attempting to optimize for locality. The instructions are then
decoded from byte sequences into instruction structures (unless exceptions are
detected). Each instruction is then translated via the XLATE and UCODE
blocks into a uop sequence which implements that instruction. This translation is
complex in and of itself; for simpler instructions, a table is consulted and a
fixed sequence of uops is generated. For more complicated instructions, a {\em
  prelude} sequence of uops will lead to a trap into {\em microcode} that has
support for conditional branching and looping controlled by UCODE.  Microcode
is generally stored in an on-chip ROM. We term the composition of these sources
of uops a {\em micro-op program} or {\em micro-program}. The uops which result
from the stepping of the micro-program are then scheduled (and possibly
reordered) for dispatch in EXEC. The source data for the uops are either
coming from registers or memory loads, and results either being written to
registers or stored to memory.

\subsection{Microcode Model}


In a fashion similar to the formal definition of {\tt x86isa}, we define a
formal model of the micro-architecture in ACL2 that we term the {\em microcode
  model} or {\em ucode model} (see Figure~\ref{fig:ucode-model}). This model is
defined by its state and an effect of the execution of a uop to the state.

The uop format in the ucode model
consists of an opcode, source and destination register identifiers, and
immediate data. There are additional fields in the uop which are used to
help optimize the scheduling and execution of the uops, which we
elide here. Note that examples of uops and micro-programs will be
presented in Sections~\ref{sec:shrd} and~\ref{sec:vpshrdq}.

The ucode model state includes a program counter ({\it PC}), several sets of registers which implement the
x86 ISA registers and flags, as well as configuration and micro-architecture specific registers
used by the uops. The {\it PC} is a structure comprised of a prelude
sequence of uops, a trap address into a fixed microcode ROM, and a set of
additional side parameters to further refine the generated uops.
The ucode state also includes a configuration object and a
simple memory which extend the programmer-visible state in {\tt x86isa}.


\begin{figure}[ht]
\begin{tabbing}
\hspace{.05 in} \= \hspace{.43 in} \= \hspace{.1 in} \= \kill
{\bf type} {\tt ucode-model-PC} {\bf is} \\
\> {\bf tuple} $\langle${\it prelude, rom-address, side-params}$\rangle$ \\
\\
{\bf type} {\tt ucode-model-state} {\bf is} \\
\> {\bf tuple} $\langle${\it PC, config, registers, memory}$\rangle$ \\
\\
{\bf func} {\tt ucode-model-step} {\it (ustate)} {\bf is} \\
\> {\it uop} \> = \> \texttt{\textbf{ucode-get-uop}}{\it (ustate.PC, ustate.config)} \\
\> {\it data} \> = \> {\tt ucode-fetch-data}{\it (uop, ustate)} \\
\> {\it results} \> = \> \texttt{\textbf{ucode-exec}}{\it (uop, data, ustate.config)} \\
\> {\it new-PC} \> = \> \texttt{\textbf{ucode-next-pc}}{\it (ustate.PC,  results)} \\
\> {\it next-ust} \> = \> {\tt ucode-update-state}{\it (results, new-PC, ustate)} \\
\> {\bf return} {\it next-ust} \\
\\
{\bf NOTE:} The functions in bold are either verified against \\RTL blocks or use symbolic simulation of RTL blocks in\\ their definition.
\end{tabbing}
\caption{Ucode Model} 
\label{fig:ucode-model}
\end{figure}


The ucode model also defines a state update function {\tt
  ucode-model-step} outlined in Figure~\ref{fig:ucode-model}. The
function {\tt ucode-get-uop} constructs the next uop defined by the {\it PC}
structure (either as a next uop in the prelude or retrieved from ROM
if {\it PC} is a rom-address). The function {\tt ucode-fetch-data} pulls the
data from registers and memory needed for the uop computation. The
function {\tt ucode-exec} defines the computation of the given uop on
the input data and returns the results of this computation (which can
also include any execute-time exceptions). The function {\tt
  ucode-next-pc} computes the next value for {\it PC} by either stepping
through the prelude or updating the rom-address, depending on whether
a branch was taken or not. A special address is used to signal a halt
or completed micro-program. Finally, the {\tt ucode-update-state}
function stores all results from the uop execution in appropriate
locations in registers and memory, as well as updates {\it PC} to {\it
  new-PC}.

The {\tt ucode-exec} function defines the operational semantics of every uop
supported in the micro\hyp{}architecture. These uops include simple integer
operations, more complex floating\hyp{}point and vector operations, memory loads and
stores, branching and jumps, along with many more. The full specification of the
uops is beyond the scope of this paper but we cover a small selection of uops
in some detail in Sections~\ref{sec:shrd} and~\ref{sec:vpshrdq}.

The definition and use of the ucode model is critical to our approach. Where
the {\tt x86isa} model provides the specification for microprocessor
verification, the ucode model provides a critical point of decomposition in the
verification effort. The {\tt ucode-exec} function is the specification for
verifying the EXEC block of the microprocessor design. In contrast,
\texttt{ucode-get-uop} and \texttt{ucode-next-pc} are defined by symbolic
execution of the corresponding UCODE and XLATE RTL blocks.
The composition of {\tt ucode-model-step} applied to the sequence of uops
generated by DECODE and XLATE and uops stored in ROM is then verified against
the {\tt x86-exec} function from {\tt x86isa}. We detail more of this in
Section~\ref{sec:overview}.

\subsection{Introduction to ACL2 and Supporting Tools} \label{sec:acl2}


All of our work is done using the ACL2 theorem prover~\cite{acl2:home} with an interface to
external SAT solvers. ACL2 is a theorem prover supporting
proofs in an untyped first-order logic, with some support for higher-order logic
styles of definition. ACL2 is written in and runs in Common Lisp environments, where
compliant functions defined in ACL2 have compiled Common Lisp counterparts
supporting efficient evaluation. In addition, ACL2 supports the intrinsic
capability of defining functions in Common Lisp that generate other
definitions and logical events. This support of macros in ACL2 is critical to
our approach since we use them to generate and prove requisite
lemmas from large data structures that codify x86 instructions and
uop sequences, and automate ucode proofs. Section~\ref{sec:automate} will cover more of these uses of model
definitions in automating proofs and building supporting tools.

In addition, we use many existing tools and libraries written in ACL2. We use
the {\tt VL} toolset~\cite{vl-github,vl-acl2-doc} to parse and translate microprocessor RTL
definitions written in SystemVerilog into syntax trees. The {\tt SV}
library~\cite{sv-github,sv-acl2-doc} takes these syntax trees from {\tt VL} and produces
semantic next-state functions for signals in the
design. Built on top of {\tt SV} is a multi-cycle extraction tool {\tt
  SVTV}\cite{svtv-doc}. {\tt SVTV} supports the generation of function definitions
in ACL2 that correspond to applying inputs to the design, stepping the design
some specified number of clock cycles, and extracting relevant outputs along the
way. For the design blocks we target, we have defined {\tt SVTV}s that capture
the effects of decoding a single instruction and executing a single instruction
or uop. This allows us to view these blocks as the corresponding ACL2 function
definitions. We use the names SV-DECODE, SV-XLATE, SV-UCODE, and SV-EXEC to
represent the ACL2 input to output mapping functions derived from these {\tt
  SVTVs}, and refer to them as the {\em SV design functions} as a group.

The {\tt GL} tool~\cite{10-swords-dissertation,11-swords-bit-blasting}
is a verified prover of ACL2 theorems on finite domains. {\tt GL} uses
symbolic simulation of ACL2 function definitions to reduce finite ACL2
theorems into propositional logic formulas. These propositional logic
formulas are then either proven with BDDs, or simplified using AIG
algorithms and transmitted to a SAT solver to check if they are true
or produce a counterexample. The BDD and AIG algorithms are written in
ACL2 and proven correct. We use external SAT solvers but the proofs
from the SAT solvers are checked for validity. We make extensive use
of {\tt GL}, from proving the necessary correspondence between the SV
design functions and the corresponding x86 and uop model functions, to
defining tools for exploration of possible proofs and generation of
constraints.

\section{Overview}\label{sec:overview}

Our focus is verifying the correct operation of the RTL-level definitions
implementing the DECODE, XLATE, UCODE, and EXEC blocks of the
microprocessor. The primary function of these blocks in the microprocessor is
the correct decode and execution of x86 instructions via correct uop execution
and sequencing---these are the highlighted blocks from
Figure~\ref{fig:microproc-org} and correspond to the bolded functions in the
x86 and ucode models. Our goal is to prove the correctness of a single
instruction invocation through these RTL blocks from a generalized legal state
with assumptions ensuring no interference or impedance to the execution of the
instruction. Importantly, the focus on a single instruction invocation allows
us to reduce the scope of the SV design functions we need for proofs, and this
in turn dramatically reduces the times spent in {\tt GL} for symbolic
simulation or SAT solving. Also, the single instruction focus reduces the need
to specify invariants, which in turn reduces the fragility of the proofs to
changes in the RTL.

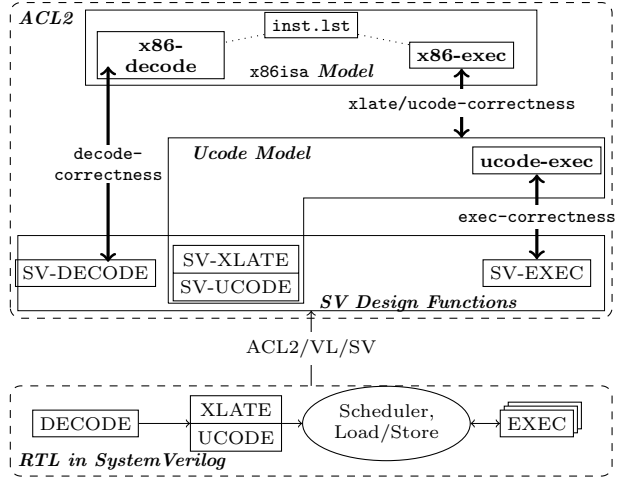
\begin{figure}[ht]
\centering
\begin{tikzpicture}

\draw   (0,1.9) [dashed,thin,rounded corners] rectangle (8,6.1);
\node[node font=\scriptsize,align=left,right] at (0,5.9) {{\bf {\em ACL2}}};

\draw   (0,-0.2) [dashed,thin,rounded corners] rectangle (8,1);
\node[node font=\scriptsize,align=left,right] at (0,0) {{\bf {\em RTL in SystemVerilog}}};
\draw	(1,0.5) node[node font=\scriptsize,align=center,draw] (rtldecode)  {DECODE}	
        (3,0.5) node[rectangle split,rectangle split parts=2,node font=\scriptsize,align=center,draw] (rtlxlate/ucode) {XLATE \nodepart{second}UCODE}
        (5,0.5) node[ellipse,node font=\scriptsize,align=center,draw] (rtlscheduler) {Scheduler,\\Load/Store}
        (7,0.5) node[double copy shadow,fill=white, node font=\scriptsize,align=center,draw] (rtlexec) {EXEC};
\path (rtldecode) edge[->, very thin] (rtlxlate/ucode);
\path (rtlxlate/ucode) edge[->, very thin] (rtlscheduler);
\path (rtlscheduler) edge[<->, very thin] (rtlexec);

\draw[->] (4,1) -- (4,2);
\node[node font=\scriptsize,align=center,fill=white] at (4,1.5) {ACL2/VL/SV};

\draw	(0.1,2) [very thin] rectangle (7.9,3);
\node[node font=\scriptsize,align=left,right] at (4,2.1) {{\bf {\em SV Design Functions}}};
\draw	(1,2.5) node[node font=\scriptsize,align=center,draw] (svdecode) {SV-DECODE}
        (3,2.5) node[rectangle split,rectangle split parts=2,node font=\scriptsize,align=center,draw] (svxlate) {SV-XLATE \nodepart{second}SV-UCODE}
        (7,2.5) node[node font=\scriptsize,align=center,draw] (svexec) {SV-EXEC};

\draw	(7,4) node[node font=\scriptsize,align=center,draw] (ucexec) {{\tt {\bf ucode-exec}}};

\path (svexec) edge[<->, very thick] node[node font=\scriptsize\tt\bfseries,align=center,fill=white]{exec-correctness} (ucexec);

\node[node font=\scriptsize,align=left,right] at (2.3,4.1) {{\bf {\em Ucode Model}}};
\draw [very thin] (2.1,2.1) -- (2.1,4.3) -- (7.9,4.3) -- (7.9,3.5) -- (3.9,3.5) -- (3.9,2.1) -- (2.1,2.1);

\draw	(1,5) [very thin] rectangle (7,6);
\node[node font=\scriptsize,align=center] at (4,5.2) {{\bf {\em {\tt x86isa} Model}}};
\draw	(2,5.4) node[node font=\scriptsize,align=center,draw,text width=1.5cm] (x86decode) {{\tt {\bf x86-decode}}}
        (4,5.8) node[node font=\scriptsize,align=center,draw] (instlst) {{\tt inst.lst}}
        (6,5.4) node[node font=\scriptsize,align=center,draw] (x86exec) {{\tt {\bf x86-exec}}};
\path (instlst) edge[-,dotted] (x86decode) edge[-,dotted] (x86exec);

\draw (1.3,5.25) edge[<->, very thick] node[node font=\scriptsize\tt\bfseries,align=center,fill=white,inner sep=1] (decodecorrectness) {decode-\\correctness} (1.3,2.65);
\path (x86exec) edge[<->, very thick] node[node font=\scriptsize\tt\bfseries,align=center,fill=white] (xlateucodecorrectness) {xlate/ucode-correctness} (6,4.3);

\end{tikzpicture}
\caption{Top-level Correctness Decomposition}
\label{fig:pretty-picture}
\end{figure}

As stated before, we define all functions and prove all theorems in the ACL2
theorem prover. 
Our goal is to prove that the SV design functions correspond to {\tt
x86isa} and ucode model functions. Our correctness statements for each
of these blocks are defined in Figure~\ref{fig:specs}, while
Figure~\ref{fig:pretty-picture} depicts how these pieces fit together
for checking single-instruction correctness.

%

\begin{figure}[ht]
\small
\begin{tabbing}
\hspace{.1 in} \= \hspace{.1 in} \= \hspace{.1 in} \= \kill
{\bf theorem} {\tt decode-correctness} {\bf is} \\
\> $\forall${\it(bytes, cfg)}$:$ \\
\> \> {\tt get-instr}(SV-DECODE({\tt map-decode}({\it bytes, cfg}))) $=$ \\
\> \> \texttt{\textbf{x86-decode}}({\it bytes, cfg}) \\
\\
{\bf theorem} {\tt xlate/ucode-correctness} {\bf is} \\
\> $\forall$({\it instr, state})$:$ \\
\> \> {\tt run-xlate/ucode}({\it instr, state}) $=$ \\
\> \> \texttt{\textbf{x86-exec}}({\it instr, state}) \\
\\
{\bf theorem} {\tt exec-correctness} {\bf is} \\
\> $\forall$({\it uop, data, cfg})$:$ \\
\> \> {\tt get-results}(SV-EXEC({\tt map-exec}({\it uop, data, cfg}))) $=$ \\
\> \> \texttt{\textbf{ucode-exec}}({\it uop, data, cfg})
\end{tabbing}
\caption{Correctness Statements for Target Blocks}
\label{fig:specs}
\end{figure}

%

The mapping functions {\tt map-decode} and {\tt map-exec} referenced in
Figure~\ref{fig:specs} build input signal bindings for the SV design
functions. The extraction functions {\tt get-instr} and {\tt get-results} pull
out specification-level results from the output signals of the SV design
functions. These {\tt map} and {\tt get} functions establish the correspondence of
runs/steps in the design and in the model. Although the proofs for {\tt
  decode-correctness} and {\tt exec-correctness} are considerable (especially
{\tt exec-correctness}, which we will discuss in more detail in
Section~\ref{sec:uops}), their statements are fairly straightforward
correspondence theorems. The property of {\tt xlate/ucode-correctness} is less
direct. The main reason for this is that the specification for the XLATE and
UCODE design blocks is complex---XLATE and UCODE produce and run a
micro-program to completion, the final results of which correspond to the
results of {\tt x86-exec}.

Before considering the definition of {\tt run-xlate/ucode}, we first return to the ucode
model and the functions which define {\tt ucode-model-step} (see Figure~\ref{fig:ucode-model}). The functions
{\tt ucode-fetch-data} and {\tt ucode-update-state} involve accessing and updating register
banks and memory, and correspond to blocks of the microprocessor we do not
target.  As such, we simply define models of these behaviors as functions in
ACL2. We also define the function {\tt ucode-exec} as a model of the EXEC block
of the microprocessor, but for each uop, we prove that the EXEC block is
consistent with {\tt ucode-exec} in {\tt exec-correctness}. For the definitions of the
functions {\tt ucode-get-uop} and {\tt ucode-next-pc}, we use the SV-UCODE design function
directly. Specifically, we split the SV-UCODE design function into
SV-UCODE-READ, which is responsible for creating uops in the design for a given PC, and
SV-UCODE-STEP, which determines the next rom-address, depending on branching. These
ucode-model functions are defined in Figure~\ref{fig:additional-ucode-model-defs}.
\begin{figure}[ht]
\small
\begin{tabbing}
\hspace{.1 in} \= \hspace{.1 in} \= \hspace{.23 in} \= \hspace{.1 in} \= \kill
{\bf func} {\tt ucode-get-uop} ({\it PC, config}) {\bf is} \\
\> {\bf if} {\tt in-prelude}{\it(PC)} {\bf then} {\tt first-prelude}{\it(PC)} \\
\> {\bf else} \\
\> \> {\tt get-uop}(SV-UCODE-READ({\tt map-uc-read}=({\it PC, config}))) \\
\\
{\bf func} {\tt ucode-next-pc}({\it PC, results}) {\bf is} \\
\> {\bf if} {\tt in-prelude}{\it(PC)} {\bf then} {\tt remove-first-prelude}{\it(PC)} \\
\> {\bf else} \\
\> \> {\tt get-pc}(SV-UCODE-STEP({\tt map-uc-step}({\it PC, results})))
\end{tabbing}
\caption{Ucode Model Using SV-UCODE functions}
\label{fig:additional-ucode-model-defs}
\end{figure}

The definition of the function {\tt run-xlate/ucode} is then provided in
Figure~\ref{fig:run-xlat-ucode}. {\tt run-xlate/ucode} first calls
SV\hyp{}XLATE (with map and get functions to transfer from specification objects to
design signals and back) to generate the initial {\it pc} for ucode execution, and
then calls {\tt run-ucode-model}, which executes {\tt ucode-model-step} until a halt
address is reached and the instruction results are extracted.

\begin{figure}[ht]
\small
\begin{tabbing}
\hspace{.1 in} \= \hspace{.1 in} \= \hspace{.23 in} \= \hspace{.1 in} \= \kill
{\bf func} {\tt run-ucode-model} ({\it ustate}) {\bf is} \\
\> {\bf if} {\tt is-halted}({\it ustate}) {\bf then} {\it ustate} \\
\> {\bf else} \\
\> \> {\tt run-ucode-model}({\tt ucode-model-step}({\it ustate})) \\
\\
{\bf func} {\tt run-xlate/ucode}({\it instr, state}) {\bf is} \\
\> {\it cfg} \> \> = {\it state.config} \\
\> {\it pc} \> \> = {\tt get-init-pc}(SV-XLATE({\tt map-xlate}({\it instr, cfg}))) \\
\> {\it ustate} \> \> = {\tt make-ucode-state}({\it pc, state}) \\
\> {\it ustate'} \> \> = {\tt run-ucode-model}({\it ustate}) \\
\> {\it results} \> \> = {\tt extract-instr-results}({\it ustate' }) \\
\> {\bf return} {\it results}
\end{tabbing}
\caption{{\tt run-xlate/ucode} function}
\label{fig:run-xlat-ucode}
\end{figure}

Our direct use of SV-UCODE and SV-XLATE design functions in defining the ucode
model {\tt run-xlate/ucode} has the significant benefit of not needing to
define and maintain a separate specification for these blocks. This is important
because the operation and even interfaces of these blocks (and especially
microcode ROM itself) are fairly complex and can change rapidly during design
iteration. This approach does introduce some challenges when we attempt
proofs. In particular, in order to automate proofs of {\tt
  xlate/ucode-correctness}, we will need to have the successive expansions of
the {\tt ucode-model-step} function not cause a significant explosion in
either the symbolic simulation of {\tt GL} or the subsequent calls to SAT
solvers. We assist this process by identifying criteria that are required of
earlier steps in {\tt GL} processing to keep subsequent steps from being too
expensive. One particularly important criteria is that we have sufficient
constraints on the instruction input to {\tt run-xlate/ucode} to ensure that
the (symbolic) PC produced by SV-XLATE has a sequence of fixed uop opcodes
(i.e., data, source, destination fields may be symbolic, but the actual uop operations
are fixed). We implement these criteria as prechecks before attempting to
submit a {\tt xlate/ucode-correctness} lemma instance.

%
We finally tie together the target specifications from Figure~\ref{fig:specs}
to state our single-instruction correctness goal in
Figure~\ref{fig:single-exec}. The function {\tt run-xlate/ucode-RTL} is the
same as {\tt run-xlate/ucode}, except that {\tt ucode-exec} is replaced with the
RTL equivalent {\tt get-results}(SV-EXEC({\tt map-exec}({\it uop, data, cfg}))) from
{\tt exec-correctness}. We state (informally) that {\tt
  single-instruction-correctness} captures the requirements for verifying
correct instruction operation in the target RTL blocks while making reasonable
(unverified) assumptions on the other RTL blocks in the microprocessor.

\begin{figure}[ht]
\small
\begin{tabbing}
\hspace{.01 in} \= $\langle$dx, instr$\rangle$\=  \hspace {0.05 in} \= \kill
{\bf func} {\tt x86-step}({\it bytes, state}) {\bf is} \\
\> $\langle${\it dx, instr}$\rangle$ \> = \>\texttt{\textbf{x86-decode}}({\it bytes, state.config}) \\
\> $\langle${\it ex, rslts}$\rangle$ \> = \>\texttt{\textbf{x86-exec}}({\it instr, state}) \\
\> {\bf return} $\langle${\it dx, ex, rslts}$\rangle$ \\
\\
{\bf func} {\tt rtl-step}({\it bytes, state}) {\bf is} \\
\> {\it cfg} \> = \> {\it state.config} \\
\> $\langle${\it dx, instr}$\rangle$ \> = \>{\tt get-instr}(SV-DECODE({\tt map-decode}({\it bytes, cfg}))) \\
\> $\langle${\it ex, rslts}$\rangle$ \> = \>{\tt run-xlate/ucode-RTL}({\it instr, state}) \\
\> {\bf return} $\langle${\it dx, ex, rslts}$\rangle$ \\
\\
{\bf theorem} {\tt single-instruction-correctness} {\bf is} \\ 
\> $\forall$({\it bytes, state})$:$ \\
\> \hspace{0.1 in}{\tt x86-step}({\it bytes, state}) $=$ {\tt rtl-step}({\it bytes, state})
\end{tabbing}
\caption{Single-Instruction Execution Specification}
\label{fig:single-exec}
\end{figure}

\subsection{Verification of Uops: {\tt exec-correctness}}\label{sec:uops}


Formal models of microprocessors usually include an effect of one step of the
machine on the state.  The granularity of the step depends on the objective of
verification. In our case, the smallest step we consider is an execution of one
uop. In most published work, operational semantics is created by authors of the
model and validated (or not) by testing. Our work differs from that approach in
the fact that our operational semantics of uops provably reflects behavior of
the underlying design, with the exception of uops which implement loads and
stores---we only model memory accesses. In particular, we prove the property
called {\tt exec-correctness} (see Figure~\ref{fig:specs}), i.e., each uop that
can be dispatched to an execution unit will produce the effects described by our
uop specification functions.  These same uop specifications are also
used in proving {\tt xlate/ucode-correctness} --- this allows us to replace a
uop's RTL implementation with these specifications while doing microcode proofs.
Figure~\ref{fig:pretty-picture} shows the central position these specification
functions occupy in our project.

The formal verification of {\tt exec-correctness} is an ongoing effort
at Centaur Technology that has been developed over several years.  This work is
usually done early on in the design cycle for catching functional bugs
in the execution units as soon as possible.  To illustrate the scale
of this effort, we proved {\tt exec-correctness} for about 600 uops
for our last project.  In these proofs, we allow multiple uops in the
pipeline, and only assume that the scheduler worked as intended,
i.e., did not schedule uops that would collide while writing results. For
instance, a uop with a 4-cycle latency cannot be scheduled to start on the
same unit one cycle after a uop with a 3-cycle latency was dispatched to the
unit.

\subsubsection{Uop Specifications}\label{sec:uops-spec}

Uops are dispatched by a scheduler to individual execution units along
with all the information that the unit may need to process them---e.g.,
opcode, source and destination sizes, operands, immediate
values, etc. At this point, the facts of where the operands came from
and where the result is heading are abstracted away.  The core of
operational semantics for every uop is a function that operates on
data types defined in the ISA: integers (signed and unsigned), packed
integers, floating point numbers, bit vectors, etc.  The results are
again those data types and (where appropriate) bit vectors
representing exceptions.

In Figure~\ref{fig:and-spec-function}, we present the specification of
a very simple integer uop, logical {\tt AND}, with flag computations
elided for presentation's sake.  We will see this uop again in
Section~\ref{sec:shrd} (specifically, in
Table~\ref{table:shrd-uops-2}) later.
\begin{figure}[ht]
\small
\begin{tabbing}
\hspace{.01 in} \= $\langle${\it operand1, operand2}$\rangle$ \=  \hspace {0.08 in} \= \kill
{\bf func} {\tt and-spec}({\it input}) {\bf is} \\
\> $\langle${\it srcsize, dstsize}$\rangle$ \> = \>\texttt{get-sizes}({\it input}) \\
\> $\langle${\it operand1, operand2}$\rangle$ \> = \>\texttt{get-operands}({\it input}) \\
\> {\it arg1} \> = {\tt truncate}({\it srcsize, operand1}) \\
\> {\it arg2} \> = {\tt truncate}({\it srcsize, operand2}) \\
\> {\it result} \> = {\tt truncate}({\it dstsize, arg1 \& arg2}) \\
\> {\bf return} {\it result}
\end{tabbing}
\caption{{\tt and-spec}: Specification Function for logical {\tt AND}}
\label{fig:and-spec-function}
\end{figure}

\subsubsection{Uop Correctness Proofs}\label{sec:uops-correctness}



At Centaur Technology, we have been doing formal verification of execution
units---those that perform logical, integer, SIMD integer, and
floating-point operations---for over a
decade~\cite{HS09,hardin-centaur,memocode11}. 
Formal methods have proven to be very effective in the validation of
data-path intensive designs. With the size of operands in the x86 ISA
growing to 512 bits, targeted or random simulations have
become inadequate to cover arithmetic and cryptographic units.  At the
same time, increasing capacity of SAT solvers and evolving symbolic
simulation methods make the verification of these units mostly
automatic. About ten percent of the uops and operations still require human assistance,
e.g., various flavors of square-root and divide, and floating-point add,
multiply, and fused multiply-add.

All uop proofs are done within the scope of the EXEC hardware block, which
includes execution units for all types of operations as sub-modules.  The
benefit of working with the entire EXEC block instead of a specific sub-module
for uop verification is that the hardware interface of EXEC changes a lot less
frequently than that of the sub-modules, which allows us to write our
specification functions in a uniform way.  The formal model of EXEC is rebuilt
automatically every time the hardware design changes, and thus, the resulting SV
design functions are always up-to-date.

The SV design functions corresponding to EXEC are amenable to various ACL2 proof
techniques, like symbolic simulation and propositional logic checks using GL.
Since almost all uops have fixed latency (i.e., the computation finishes in a
fixed number of clock cycles), symbolic simulation is our main tool to prove
their correctness. GL symbolically executes the given module, extracts results
and flags, and compares them to the corresponding specification function. In
most cases, this happens without any user intervention.  If a function is too
complex to be processed by GL directly, the user can split it into cases (e.g.,
for floating-point adders), or SV-EXEC can be decomposed into several runs of
the sub-units (e.g., for multipliers). For both these cases, the decomposition
is proven correct using ACL2. For cases like dividers that have variable
latency, more sophisticated proof techniques are needed---e.g., the user has to
guide ACL2 by discovering and proving inductive invariants.

We present what {\tt exec-correctness} looks like for a simple uop, logical {\tt
  AND}, in Figure~\ref{fig:logical-and-correctness-thm} where {\tt and-spec} is
the specification of {\tt AND} in {\tt ucode-exec}.

\begin{figure}[ht]
\small
\begin{tabbing}
\hspace{.1 in} \= \hspace{.1 in} \= \hspace{.1 in} \= \kill
{\bf theorem} {\tt and-exec-correctness} {\bf is} \\
\> $\forall$({\it input}): \\
\> \> {\tt get-results}(SV-EXEC({\tt map-exec}({\tt AND}, {\it input}))) $=$ \\
\> \> \texttt{\texttt{and-spec}}({\it input})
\end{tabbing}
\caption{Logical {\tt AND} Correctness Theorem}
\label{fig:logical-and-correctness-thm}
\end{figure}


As a part of establishing {\tt exec-correctness}, our proof scripts
automatically generate documentation in HTML about the proof's details and
coverage.  This way, the documentation is always in sync with the proofs, and
accessible to hardware designers.


\section{Illustrative Example: Verifying {\tt SHRD}}\label{sec:shrd}

In this section, we describe in detail how we verify the
implementation of the instruction {\tt SHRD}, which stands for
``shift-right double''.  In the following section
(Section~\ref{sec:vpshrdq}), we briefly describe {\tt VPSHRDQ}, which
is another instruction that performs ``shift-right double'' but on
{\em packed data}, to illustrate how we handle more
complicated instructions.

{\tt SHRD}, along with its shift-left counterpart {\tt SHLD}, was
introduced to aid bit string operations, and is supported on all Intel
processors since the 80386.  These instructions manipulate
general-purpose x86 registers, memory (depending on the variant used),
and the rflags register.

\subsection{Instruction Specification}\label{sec:shrd-inst-spec}
The {\tt SHRD} instruction takes three operands:
\begin{Verbatim}[fontsize=\small]
SHRD <destination>, <source>, <shiftAmt>
\end{Verbatim}
The x86 ISA provides two variants of this instruction: in one, the
{\tt shiftAmt} is an 8-bit immediate and in the other, it is the 8-bit
wide {\tt CL} register.  For both of these variants, the {\tt
  destination} can either be a register or memory operand, the source
must be a register, and the {\tt destination} and {\tt source} must be
of the same size: either 16, 32, or 64 bits.

This instruction behaves as follows: the {\tt destination} is shifted right by
a value indicated by {\tt shiftAmt} (which is masked appropriately, as dictated
by the instruction size) and the resulting empty bit positions are filled with
bits shifted in from the {\tt source} (least-significant bit first).  {\tt
  SHRD} can also affect flags; though we do specify and verify flag
computations, we omit flag-related discussions for the sake of brevity.  The
specification function of {\tt SHRD} from {\tt x86-exec} in our {\tt x86isa}
model is called {\tt shrd-spec}. This function describes the core operation
of {\tt SHRD}, without dealing with machine aspects like operand addressing and
fetching.

In this paper, we focus on the following {\tt SHRD} variant:
\begin{Verbatim}[fontsize=\small,commandchars=\\\{\}]
variant: SHRD RCX, RDX, <imm8>
bytes:   0x48 0x0F 0xAC 0xD1 <imm8>
\end{Verbatim}
That is, it is a 64-bit {\tt SHRD} instruction whose destination is a
register and which takes an immediate byte as the {\tt shiftAmt}
operand.  See Figure~\ref{fig:shrd-instruction} which shows a concrete
run of this variant.
\begin{figure}[ht]
\centering
\begin{tikzpicture}
 \node at (4,1.75) {{\footnotesize{\em --- Initial Values  --- }}};

  \draw (0,1) [thick] rectangle (4,1.5);
  \node at (2,1.25) {{\scriptsize{\tt RDX := 0x1122\_3344\_5566\_7788}}};
  \draw (4,1) [thick] rectangle (8,1.5);
  \node at (6,1.25) {{\scriptsize{\tt RCX := 0x0123\_4567\_89AB\_CDEF}}};

  \node at (4,0.75) {{\footnotesize{\em --- Final Values  --- }}};

  \draw (0,0) [thick] rectangle (4,0.5);
  \node at (2,0.25) {{\scriptsize{\tt RDX := 0x1122\_3344\_5566\_7788}}};
  \draw (4,0) [thick] rectangle (8,0.5);
  \node at (6,0.25) {{\scriptsize{\tt RCX := 0x7788\_0123\_4567\_89AB}}};

\end{tikzpicture}
\caption{{\tt SHRD RCX, RDX, 16}}
\label{fig:shrd-instruction}%
\end{figure}

\subsection{Microcode Implementation}\label{sec:shrd-ucode-impl}
On one of Centaur Technology's processor designs, the variant of {\tt SHRD}
we are interested in has two prelude uops,
followed by a trap to a routine in the microcode ROM labeled {\tt
  ent\_shrdEvGv\_64reg}.  The microcode ROM contains a compact representation
of the uops in order to conserve space.  As such, there is an additional step
required to obtain the uops corresponding to a ROM instruction---the {\em
  microsequencer} hardware block (the part of UCODE defining SV-UCODE-READ) is
responsible for translating this compact representation to a uop sequence.  The
microprogram corresponding to this variant is presented in
Tables~\ref{table:shrd-uops-1} and ~\ref{table:shrd-uops-2}; uops that compute
only the flag values are elided.  In the second column of these tables, we show
the contents of the relevant state components for the concrete run shown in
Figure~\ref{fig:shrd-instruction}.
\begin{table}[ht]
\centering
  \caption{{\bf {\tt SHRD}: Prelude Uops}}
  \label{table:shrd-uops-1}
  \vspace{-10pt}
  {\small
   \begin{tabular}{p{3cm} | p{5.5cm}}
     \hline
   {\bf Uop} & {\bf Concrete Run \& Description} \\
   \hline
   {\tt MOVSX G2, RCX \newline{\footnotesize(SSZ: 64; DSZ: 64)}} &
   {\tt G2 $\leftarrow$ 0x0123\_4567\_89AB\_CDEF}\newline {\footnotesize {\it Move {\tt RCX} to internal register {\tt G2}}} \\
   \hline
   {\tt MOVZX G3, <imm8> \newline{\footnotesize(SSZ: 8; DSZ: 64)}} &
   {\tt G3 $\leftarrow$  16} \newline{\footnotesize {\it Move immediate to internal register {\tt G3}}} \\
   \hline
 \end{tabular}}
\caption{{\bf {\tt SHRD}: Uops in {\tt ent\_shrdEvGv\_64reg}}}
\label{table:shrd-uops-2}
\vspace{-10pt}
{\small
 \begin{tabular}{p{3cm} | p{5.5cm}}
 \hline
 {\tt AND G3, G3, 63 \newline{\footnotesize (SSZ: 8; DSZ: 64)}} &
 {\tt G3 $\leftarrow$  16} \newline {\footnotesize {\it Mask immediate operand}} \\
 \hline

 {\tt MOV G10, -1 \newline{\footnotesize(SSZ: 64; DSZ: 64)}} &
 {\tt G10 $\leftarrow$  0xFFFF\_FFFF\_FFFF\_FFFF \newline{\footnotesize {\it Move -1 to internal register {\tt G10}}}} \\
 \hline

 {\tt JE G3, 0, ent\_nop \newline{\footnotesize(SSZ: 16; DSZ: 16)}} &
 {\tt No jump taken} \newline{\footnotesize {\it Jump to routine {\tt ent\_nop} if {\tt G3} == 0}} \\
 \hline

 {\tt SUB G5, 0, G3 \newline{\footnotesize(SSZ: 32; DSZ: 32)}} &
 {\tt G5 $\leftarrow$  0xFFFF\_FFF0; ZF $\leftarrow$ 0 } \newline{\footnotesize {\it Store {\tt -G3} in internal register {\tt G5}; clear the zero flag because result is non-zero}} \\
 \hline

 {\tt SHR{\tiny <!ZF>} G10, G10, G5 \newline{\footnotesize(SSZ: 64; DSZ: 64)}} &
 {\tt G10 $\leftarrow$  0xFFFF } \newline{\footnotesize {\it Shift {\tt G10} right by {\tt (G5 \& 63)} if {\tt ZF} == 0}} \\
 \hline

 {\tt AND{\tiny <ZF>} G10, G10, 0 \newline{\footnotesize(SSZ: 64; DSZ: 64)}} &
 {\tt G10 $\leftarrow$  0xFFFF } \newline{\footnotesize {\it Set {\tt G10} to 0 if {\tt ZF} == 1}} \\
 \hline

 {\tt AND G6, RDX, G10 \newline{\footnotesize(SSZ: 64; DSZ: 64)}} &
 {\tt G6 $\leftarrow$  0x7788 } \newline{\footnotesize {\it Store {\tt (RDX \&  G10)} in internal register {\tt G6}}} \\
 \hline

 {\tt SHR G7, G2, G3 \newline{\footnotesize(SSZ: 64; DSZ: 64)}} &
 {\tt G7 $\leftarrow$  0x0000\_0123\_4567\_89AB } \newline{\footnotesize {\it Store {\tt (G2 >> G3)} in {\tt G7}}} \\
 \hline

 {\tt SHL G2, G7, G3 \newline{\footnotesize(SSZ: 64; DSZ: 64)}} &
 {\tt G2 $\leftarrow$  0x0123\_4567\_89AB\_0000 } \newline{\footnotesize {\it Store {\tt (G7 << G3)} in {\tt G2}}} \\
 \hline

 {\tt OR G2, G2, G6 \newline{\footnotesize(SSZ: 64; DSZ: 64)}} &
 {\tt G2 $\leftarrow$  0x0123\_4567\_89AB\_7788 } \newline{\footnotesize {\it Store {\tt (G2 | G6)} in {\tt G2}}} \\
 \hline

 {\tt ROR G7, G2, G3 \newline{\footnotesize(SSZ: 64; DSZ: 64)}} &
 {\tt G7 $\leftarrow$  0x7788\_0123\_4567\_89AB} \newline{\footnotesize {\it Rotate {\tt G2} right by {\tt G3} and store result in {\tt G7}}} \\
 \hline

 {\tt OR RCX, G7, G7  \newline{\footnotesize(SSZ: 64; DSZ: 64)}} &
 {\tt RCX $\leftarrow$  0x7788\_0123\_4567\_89AB } \newline{\footnotesize {\it Store the result of {\tt G7 | G7} in {\tt RCX}}} \\
 \hline

 \end{tabular}}
\end{table}

Most ucode programs are written somewhat differently from software programs
that would implement the same behavior.  For instance, in
Table~\ref{table:shrd-uops-2}, the final result is moved from the internal
register {\tt G7} to {\tt RCX} using a {\em logical or} operation, instead of a
more ``natural'' {\em move} operation.  Such choices are deliberate---microcode
programmers carefully pick uops that either have a lower latency or that reduce
code and data dependencies to aid re-ordering. The code is also written in such a way so it
can be re-used for several variants of an instruction, e.g., with different
sizes of immediate field.  Also, adding new uops is non-trivial, so
pre-existing uops have to be creatively used to implement new instructions.
The upshot is that all of this definitely needs functional verification!

\subsection{Correctness Proof}\label{sec:shrd-proof}

As described in Section~\ref{sec:motivation}, establishing an instruction's
correctness involves proving that the DECODE, XLATE/UCODE, and EXEC hardware
operate correctly. We discuss the first two for our {\tt SHRD} example below; a
general discussion of EXEC verification is in Section~\ref{sec:uops}. First,
we cover a critical data structure we use in both the {\tt decode-correctness} and
{\tt xlate/ucode-correctness} proofs.

\subsubsection{{\tt inst.lst} Data Structure}
A key part of our framework is {\tt inst.lst}~\cite{spISA2019}, which
is a data structure that we defined in the {\tt x86isa} model that
contains a listing of x86 instructions supported (or slated to be
supported) by x86 processors, including all the information needed to
decode and dispatch these instructions.  The entry for {\tt SHRD} is
depicted in Figure~\ref{fig:shrd-inst-lst}. {\tt Operands} describes
the three operands of {\tt SHRD}; for instance, the first operand is
obtained from the reg field of the ModR/M byte of {\tt SHRD}'s byte
sequence\footnote{Though not explicitly listed here, we also account
  for {\em field extensions}.  For instance, the REX.R bit (if
  present) is used along with the 3-bit ModR/M.reg field when 4 bits
  are needed for operand addressing.}, and it can either be a
general-purpose register or a memory operand. {\tt Exceptions} lists
the exceptions that {\tt SHRD} can throw, either at decode- or
execute-time.
\begin{figure}
\begin{Verbatim}[fontsize=\small,commandchars=\\\{\}]
Mnemonic:   SHRD   Opcode:   0x0F_AC
Operands:   OP1 := [ModR/M.r/m GPR MEM]
            OP2 := [ModR/M.reg GPR]
            OP3 := [IMM8]
Exceptions: #UD:    \normalfont{if LOCK prefix used}
            #GP(0): \normalfont{if the memory address is non-canonical}
            ...
\end{Verbatim}
\caption{Entry for {\tt SHRD} in {\tt inst.lst}}
\label{fig:shrd-inst-lst}
\end{figure}

\subsubsection{DECODE}\label{sec:shrd-decode-proof}
The proof of {\tt decode-correctness} for any instruction (including {\tt SHRD}) entails
showing that the incoming bytes are parsed and mapped to an appropriate and
valid instruction data structure or the correct exception is returned. The
{\tt x86-decode} specification function defines these behaviors and we split 
{\tt decode-correctness} into checking these two pieces.

For the parsing of the incoming bytes, {\tt x86-decode} pulls apart the byte sequence
to determine the proper components and fills in the instruction data structure
accordingly (assuming that the maximum limit of 15 bytes is not exceeded). The
{\tt inst.lst} structure is consulted during this parsing to determine if
certain bytes (e.g., the ModR/M or immediate data) are expected.

In addition to byte parsing, we also check if exceptions are thrown when the
instruction is illegal. An exception may be thrown for a variety of reasons,
from simple parsing violations (e.g., more than 15 bytes) to instructions which
are illegal in the current state or configuration (e.g., an instruction which
requires a certain priority level). Beyond some basic parsing exceptions, the
exception checks are defined in the {\tt inst.lst} structure. The {\tt
  x86-decode} function filters out these exceptions to only include exceptions
which are checked at decode time, as some exceptions are checked during UCODE,
or during EXEC, or in a block which we do not currently target (e.g., page
faults). For {\tt SHRD}, only one exception specifier from {\tt inst.lst} makes
it through this filter: the exception which fires if a LOCK prefix exists.
Additionally, we generate and perform a check which ensures
that any byte sequence that does not map to any
entry in {\tt inst.lst} will lead to an illegal instruction exception being
thrown.

After case splitting on opcode value and a few internal parameters of the
DECODE block, the resulting generated lemmas go through processing of {\tt GL} and
SAT solving in a range of 5 to 10 seconds. This case split produces a few
thousand cases to check across all opcodes, but these can be checked in
parallel across multiple machines.


\subsubsection{XLATE/UCODE}\label{sec:shrd-composition-proof}
After verifying that our hardware throws decode-time exceptions for
illegal byte sequences of {\tt SHRD} when the {\tt x86isa} model
mandates it should (and no such exceptions in all other cases), we
need to prove that legal byte sequences of {\tt SHRD} perform the
intended operation on our hardware. To this end, we have developed a
framework that enables us to pick and automatically populate the
candidate instruction for verification.  By using SV-DECODE, we ensure
that this population is done with provably legal values (concrete
and/or symbolic) corresponding to this candidate---that is, these
values are known to not cause any decode-time exceptions.  These
values are then passed to XLATE/UCODE, and thus, we obtain a
verification target for these blocks that is consistent with the
candidate instruction.  This process is described below using {\tt
  SHRD}, our running example.

We provide the following input to this framework in order to pick the
relevant variant of {\tt SHRD}:
\begin{Verbatim}[fontsize=\small,commandchars=\\\{\}]
Mnemonic:   SHRD   Opcode:   0x0F_AC
Variant:    Size := 64; OP1 := GPR
Mode:       64-bit mode
Indices:    OP1 := RCX; OP2 := RDX
Symbolic:   OP3
\end{Verbatim}
The first line is used to find the appropriate entry for {\tt SHRD} in
{\tt inst.lst}, which gives us information about the arity and kinds
of operands of this instruction.  The second line lets us pick the
right variant by specifying the operation width, 64, and that the
first operand should be a register, not a memory location.  The third
line picks the machine configuration, and the fourth line picks the
registers for the first two operands---note that the registers'
indices are fixed, not their contents.  The last line instructs the
framework to pick a symbolic value for the third operand, i.e., the
immediate byte that specifies the shift amount.  All of this
information is used to partially populate the instruction-related data
structures corresponding to those used in the RTL.  We then submit the
following conjecture to be bit-blasted in {\tt GL} using a SAT
decision procedure:
\begin{displayquote}
  When SV\hyp{}DECODE is given these
partially-filled instruction structures as input, an exception is
detected.
\end{displayquote}
If this conjecture is indeed a theorem, then likely there is some
contradiction in the way the framework is instructed to pick the
variant and one would need to revisit that.  Otherwise, {\tt GL}/SAT
will produce a counterexample\footnote{Partial population of the instruction 
  structure makes the SAT problem tractable.};
i.e., concrete assignments to variables in the partially-filled
instruction structures that do {\em not} lead to any exceptions. This
gives us any assignments that were missed during the partial
population of the structures.  We then generalize this counterexample
to ensure that we symbolize those parts of the instruction variant
that we care about (e.g., the immediate operand for this {\tt SHRD}
variant).  In this manner, we obtain appropriately populated
instruction structures that are consistent with our candidate {\tt
  SHRD} variant.  Typically, this entire process takes around 10
seconds, even for instructions with a more complicated encoding, like
AVX512 instructions.

These legal structures are then passed through our ACL2 model of the
XLATE and UCODE blocks (SV-XLATE and SV-UCODE), which gives us the
micro-op program--that is, the prelude uops (i.e., uops generated by
the translator) and if there is a trap to the microcode ROM, the
address of the ucode routine. As mentioned in
Section~\ref{sec:overview}, it is important for the uops' opcodes to
be fixed, though they can contain symbolic data (like the immediate
byte in our {\tt SHRD} example), and even source/destination indices
(which are fixed to {\tt RCX} and {\tt RDX} in this example).

We then attempt to prove that the single-instruction correctness
property holds for all relevant executions of this micro-op program.
That is, the effects produced by the function {\tt shrd-spec} on the
ISA-visible components of the ucode state are the same as those
produced by the implementation (i.e., uops' execution), provided that
the arguments of {\tt shrd-spec} correspond to the instruction's
operands.  These kinds of proofs can be done by techniques like the
clock functions and step-wise invariants
approaches~\cite{boyer1996mechanized,raymoore,ray-mechanical}, and
decompilation-into-logic~\cite{myreen-diff-arch,codewalker-books}, all
of which could employ either {\tt GL}/SAT's automatic bit-blasting
and/or ACL2 rewriting.  The central idea though is symbolic simulation
of these uops on our ucode model.  For the prelude uops, the ucode
model's update function directly dispatches control to the appropriate
semantic functions.  For the ucode routine, the update function dispatches
control to our microsequencer model, which translates the ROM
representation into uops; the resulting uops are then simulated in a
manner similar to the prelude uops.

It is important to mention that being able to get the prelude uops and
the trap address automatically in this manner enables us to keep up
with the constantly-changing RTL.  For non-major ``everyday'' changes
in the RTL (e.g., if the ROM addresses change or if the uops use
different internal registers), our proofs usually work without any
intervention on our part.

Note that a single microcode routine is often used to implement many
different variants of the same (or a different but similar)
instruction.  For instance, {\tt ent\_shrdEvGv\_64reg} is also used to
implement the {\tt SHRD} variant which uses the {\tt CL} register for
{\tt shiftAmt}.  Effort duplication in single-instruction verification
can be avoided by proving the sub-routine correct once for all the
variants that use it.  However, there is some merit to not decomposing
the proof in this manner; such a decomposition would involve specifying the
preconditions for the routine's correctness, which may change if the
RTL design changes.  Therefore, the decision to do proof decomposition
can be taken on a case-by-case basis---we typically do not perform
decomposition for instructions whose proofs can be automated
efficiently using bit-blasting procedures, and we typically perform
decomposition for instructions whose microcode routines need some
interactive theorem proving.

\section{AVX512 Example: Verifying \texttt{VPSHRDQ}}\label{sec:vpshrdq}

The {\tt VPSHRDQ} instruction is an AVX512 instruction intended to be
a part of future Intel processors~\cite{intel-futures-manual} (as of
this writing).  It operates on {\tt XMM}/{\tt YMM}/{\tt ZMM} registers
and memory (depending on the variant used), and does not affect any
flags.

\subsection{Instruction Specification}\label{sec:vpshrdq-inst-spec}

Intel's AVX512 instructions are their most recent SIMD (Single
Instruction on Multiple Data) instructions---i.e., the same
operation is performed on multiple data elements. {\tt VPSHRDQ} operates on 64-bit elements, and therefore, can be
thought of as the vector version of the 64-bit {\tt SHRD}.  However,
{\tt VPSHRDQ} is a much more complicated instruction than {\tt SHRD},
in part due to its more complex instruction encoding and exception
checking, and a wider data path.  Additionally, AVX512 instructions
have some features that are not supported by general-purpose
instructions.  For instance, by default, all data elements in the
operands are processed, but one can specify an {\em opmask register}
that allows their conditional processing.  That is, operation for data
element {\tt n} is performed only if bit {\tt n} of the opmask
register is set.  If this bit is not set, then either the {\tt n}-th
element of the destination retains its original value ({\em
  merge-masking}; default) or it is zeroed out ({\em zero-masking}),
based on the masking mode specified in the instruction.

The {\tt VPSHRDQ} instruction takes four operands:
\begin{Verbatim}[fontsize=\small]
VPSHRDQ <destination>[<Opmsk>][<MaskingMode>],
	<source1>, <source2>, <ShiftAmt>
\end{Verbatim}
Depending on the {\tt <Opmsk>}, the shift-right double operation is
performed on 64-bit elements in {\tt <source1>} and {\tt <source2>}
(akin to the first two operands of {\tt SHRD}), and the result,
depending on {\tt <MaskingMode>}, is either merge- or zero-masked, and
stored in the {\tt <destination>} operand.

We focus on the 512-bit variant of {\tt VPSHRDQ} that has {\tt ZMM}
registers as the first three operands:
\begin{Verbatim}[fontsize=\small,commandchars=\\\{\}]
variant: VPSHRDQ ZMM1<OpmskReg><MaskingMode>,
		 ZMM2, ZMM3, <imm8>
bytes: 0x62 0xF3 0xED 0b<MaskingMode>1001<OpmskReg>
       0x73 0xCB <imm8>
\end{Verbatim}
Like in {\tt SHRD}, the immediate byte is symbolic, and additionally,
we symbolize the opmask register and the masking mode.  This allows us
to reason about multiple invocations of the same instruction variant
in one proof---of course, we still require that this variant
translates to a fixed set of uop operations, for reasons previously described in
Section~\ref{sec:overview}.

\subsection{Microcode Implementation}\label{sec:vpshrdq-ucode-impl}
On one of our processors, the 512-bit version of {\tt VPSHRDQ} is
implemented by performing the requisite operations separately on the
low and high 256 bits of the data and then combining the results.
This variant has five prelude uops and then it traps to a microcode
ROM routine labeled {\tt avx\_double\_shift\_or\_q}; the
microprogram is listed in Tables~\ref{table:vpshrdq-uops-1}
and~\ref{table:vpshrdq-uops-2}.  Note that though {\tt VPSHRDQ} has
fewer uops than {\tt SHRD}, the behavior of each of its uops is
significantly more complicated than those of {\tt SHRD}.

\begin{table}[ht]
  \centering
  \caption{{\bf {\tt VPSHRDQ}: Prelude Uops}}
  \label{table:vpshrdq-uops-1}
  \vspace{-10pt}
  {\small
   \begin{tabular}{p{4cm} | p{5.5cm}}
   \hline
   {\bf Uop} & {\bf Description} \\
   \hline
   {\tt DLSHFTCNT T26, <IMM8> \newline{\footnotesize(SSZ: 256; DSZ: 256)}} &
   {\footnotesize {\it Store the left shift count (for operand
       {\tt ZMM3}) in internal register {\tt T26}}} \\
   \hline
   {\tt PSRLQ T27, ZMM2L, <IMM8> \newline{\footnotesize(SSZ: 256, DSZ:256)}} &
   {\footnotesize {\it Store {\tt packed-right-shift} result for the low
       256 bits of {\tt ZMM2} in internal register {\tt T27}}} \\
   \hline
   {\tt PSRLQ T29, ZMM2H, <IMM8>  \newline{\footnotesize(SSZ: 256, DSZ:256)}} &
   {\footnotesize {\it Store the {\tt packed-right-shift} result for the high
       256 bits of {\tt ZMM2} in internal register {\tt T29}}} \\
   \hline
   {\tt  PSLLVQ T28, ZMM3L, T26 \newline{\footnotesize(SSZ: 256, DSZ: 256)}} &
   {\footnotesize {\it Store the {\tt packed-left-shift} result for the low
       256 bits of {\tt ZMM3} in internal register {\tt T28}}} \\
   \hline
   {\tt  PSLLVQ T30, ZMM3H, T26 \newline{\footnotesize(SSZ: 256, DSZ: 256)}} &
   {\footnotesize {\it Store the {\tt packed-left-shift} result for the high
       256 bits of {\tt ZMM3} in internal register {\tt T30}}} \\
   \hline
 \end{tabular}}
\caption{{\bf {\tt VPSHRDQ}: Uops in {\tt avx\_double\_shift\_or\_q}}}
\label{table:vpshrdq-uops-2}
\vspace{-10pt}
{\small
   \begin{tabular}{p{4cm} | p{5.5cm}}
   \hline
   {\tt PORQ ZMM1L, T27, T28, <MaskMode>, <Opmsk> \newline{\footnotesize(SSZ: 256, DSZ:256)}} &
   {\footnotesize {\it Store {\tt (T27 packed-or T28)} in low 256 bits
of {\tt ZMM1} }} \\
   {\tt PORQ ZMM1H, T29, T30, <MaskMode>, <Opmsk> \newline{\footnotesize(SSZ: 256, DSZ:256)}} &
   {\footnotesize {\it Store {\tt (T29 packed-or T30)} in high 256
bits of {\tt ZMM1}}} \\
   \hline
   \end{tabular}}
\end{table}

\subsection{Correctness Proof}\label{sec:vpshrdq-proof}
Our verification approach, previously illustrated using {\tt SHRD}, works well
even for complex AVX512 instructions like {\tt VPSHRDQ}.  The verification of
{\tt VPSHRDQ}'s XLATE/UCODE part is very similar to that for {\tt SHRD}
(Section~\ref{sec:shrd-composition-proof}), so we omit a description here.
However, the {\tt decode-correctness} proofs of {\tt SHRD} and {\tt VPSHRDQ} do
differ (primarily in exception checking).  Another difference between the
proofs of {\tt SHRD} and {\tt VPSHRDQ} lies in the unit-level verification of
the uops themselves (Section~\ref{sec:uops})---since {\tt VPSHRDQ}'s uops are
more complicated to implement in the {\tt EXEC} RTL, the {\tt exec-correctness}
proofs were also more involved.

\subsubsection{DECODE}\label{sec:vpshrdq-decode-proof}
The {\tt decode-correctness} proof for {\tt VPSHRDQ} is again broken into two pieces:
correctness of the parsing of the incoming bytes and the exceptions
thrown. The check that parsing is correct is generated and checked in the same
way as it is for {\tt SHRD}---the incoming bytes are split based on which
prefixes, opcode bytes, and postfix bytes are included in the instruction. The
generated checks cover all variants of {\tt VPSHRDQ}.
%

Similar to {\tt SHRD}, the exception checks for {\tt VPSHRDQ} are also derived
from the {\tt inst.lst} entry for {\tt VPSHRDQ}. But, in the case of EVEX
instructions, the exception checking is more extensive. There are several
EVEX ``type-checks'' ensuring that proper values are set for the EVEX instruction
fields. For example, {\tt VPSHRDQ} uses the {\tt <Opmsk>} and the {\tt
  <MaskMode>}, but this means it cannot have an {\tt <Opmsk>} of $0$ (which
represents no mask) with the {\tt <MaskMode>} set to zero-masking. The ``types'' of the
instructions are defined in the {\tt inst.lst} structure, and the appropriate
checks are codified in {\tt x86-decode}. The resulting lemma proofs in {\tt GL} and SAT are
still quickly proven per case within a few seconds.

\section{Debugging and Automation}\label{sec:automate}

All of our formal models---RTL, ucode, and {\tt x86isa}---and specification
functions are directly executable in ACL2.  This affords us the capability to
run concrete tests in the exact same framework in which we do our proofs.  This
is useful to not only validate our models efficiently, but also to run
preliminary tests before we embark on verifying the implementation of an
instruction.  If any of our proofs fail, we can examine the failure by
executing the offending case on our models immediately---this way, we can
quickly weed out any potential failures due to, say, an incomplete
specification.  Additionally, if a proof succeeded under certain conditions but
not others, we can provide useful feedback to the logic designers by giving
them concrete ISA-level test cases that are representative of successful and
unsuccessful runs.  In case of EXEC, we can generate waveforms that exhibit
behaviors that are interesting to the logic designers.

As already discussed in Section~\ref{sec:uops-correctness}, we have been
automating unit-level uop proofs at Centaur Technology for over a decade now.  Proving
{\tt exec-correctness} for most of the uops is completely automatic now using
GL.  For {\tt decode\hyp{}correctness} proofs, we automatically generate a set of
lemmas which cover {\tt decode-correctness} for each possible opcode value
(there are roughly a thousand legal opcode values). We check that all
ill-formed incoming byte sequences cause an exception and all well-formed byte
sequences will emit a valid opcode. For each opcode, we generate checks that
correspond to each entry in the {\tt inst.lst} data structure.  We also
generate a check that for any valid opcode, if no such entry matches the
incoming byte sequences, then an exception is generated.

An example of the benefits of automation that are afforded to us by our use of
ACL2 and {\tt GL} was discussed earlier in this paper--we need to have a fixed
uop opcode sequence in the generated PC descriptor from SV-XLATE. This property
is one of several that we have diagnosed as a prerequisite for ensuring that
the subsequent {\tt GL} checks will be efficient. In our framework, we generate
a partially-symbolic PC for a given instruction specification from the user. We
not only check that the uop opcode sequence is fixed in the generated PC
resulting from SV-XLATE, but we also generate a check to ensure that no other
uop opcode sequence is possible for the given instruction specification. In
this case, we use a counterexample from a {\tt GL} check to partially fill-in a
PC value and then generate a second {\tt GL} check to ensure that this PC is
not over-constrained for subsequent proof attempts.

The scale of our problem necessitated the development of utilities that
automate many of the tasks common to every instruction; ACL2's macros are
indispensable in this respect.  We discussed an example of this in
Section~\ref{sec:shrd-composition-proof}, where we described our framework that
picks the right candidate instruction and automatically populates legal
instruction structures. Another example is the use of macros to generate lemmas for decoding ROM instructions
during ucode proofs. Even the generation of most of our proof statements is
automated.  For instance, we have utilities that generate the {\tt
  xlate/ucode-correctness} statement for AVX512 instructions.  Many instruction
implementations are amenable to automated verification using {\tt GL}/SAT
bit-blasting (e.g., those that have no loops in their microcode routines).
Automating the proof of {\tt xlate/ucode-correctness} involves careful
orchestration of the rewrite mechanism by using different theories in different
phases of the proof.


\section{Related Work}\label{sec:related}

Formal verification of microprocessors is an active area which has a long history
(since the early 1970s), both in academia and industry. There is an extensive
list of previous efforts in microprocessor verification (in x86 or otherwise),
but most are not directly applicable here, given the scope of what we address in both
the scale of industrial processor design and/or the particular requirements of
implementing the x86 ISA. As such, we mention only those works that have goals
and/or approaches that are similar to ours and that focus on industrial
processors.

The most recent report of mainstream commercial use of formal analysis
techniques to verify processor designs described how pipeline control
verification was done using bounded model checking on ARM
processors~\cite{conf/cav/Reid16}; the formal verification framework
here could also make use of data-path verification results obtained
using other formal techniques.  This work is impressive, but both its
focus and approach are different from ours.  There has been a
significant amount of work done to verify a processor's core execution
units~\cite{aagaard2000methodology,intel-i7-fv,memocode11}. Symbolic
simulation has been used to analyze microcode at Intel since
2003~\cite{aagaard2003hazards,Franzen:2010:ASS:1998496.1998520,arons2005formal}.
Their formal tool, MicroFormal, is based on SAT/SMT, and was used to
verify assertions and perform backward compatibility checks of
microcode.  Work has also been done to verify whether an x86
instruction decoder marks the boundaries of instructions
correctly~\cite{Aagaard:1998:CTP:277044.277189,aagaard1999formal,jones2002symbolic};
however, this work did not account for checking decode-time
exceptions.  Further, the x86 ISA is far more extensive and complex
now.

In contrast to these approaches, we extended our prior work on
the execution unit~\cite{HS09,hardin-centaur} as well as
microcode~\cite{centaur-itp14} verification, and along with proofs
about instruction decoding and composition of uops, are able to prove
instruction implementations correct.  Notably, the operational
semantics of uops provably reflects behavior of the RTL design.  Other
complex parts of processors have also been formally verified; e.g.,
register renaming and BUS recycle logic~\cite{kaivola2005formal},
memory systems~\cite{stewart2014processor}, etc.  Though we do not
focus on such hardware blocks yet, we plan to tackle them in the
future.

\section{Conclusions}\label{sec:future}

In this paper, we presented our progress towards comprehensive formal
verification of x86 processors designed at Centaur Technology.  We built upon
our previous work on unit-level uop verification, enhanced it,
developed a framework to verify the DECODE, XLATE/UCODE, and EXEC
blocks, and composed all these critical pieces together to prove
instruction implementations correct.

The approach presented in this paper is capable of finding bugs in the
decode, translate, microsequencer, and execution units, in addition to
microcode programs and the assembler that translates ROM
representations to uops.  Though we have not been working on this
project for long, we have already found
some bugs that could not have been found by unit-level uop
verification alone.  For example, we found instances where uops had
``don't-care'' as a source when they should have had the concrete
value 0.  We also found a case when one of many variants of an AVX512
instruction did not take the opmask register into account.  Another
example was that some AVX512 instructions did not raise a decode-time
exception when expected (e.g., some instructions throw an exception if
the non-default masking mode is chosen).  There are bugs that our
approach can indeed miss---e.g., in the memory system (load/store
uops, caches), scheduler (operand dependency and forwarding), register
mapping, pipelining (interactions among instructions), etc.

Our approach affords us the ability to adopt a divide-and-conquer
strategy for verifying instruction implementations.  Unit-level uop
verification and DECODE proofs can be done independently of each
other.  Similarly, we do not need to have the instruction and/or uop
specifications beforehand---as soon as we add an instruction's
information to {\tt inst.lst}, we can start verifying the DECODE part.
All these proofs use the same formalisms; thus, we avoid any potential
inconsistencies that may arise due to translation between different
tools/representations.  Moreover, it is straightforward to compose the
individual lemmas into a top-level theorem.

A key advantage of our approach is that we do not need to write specifications
for (or even understand) how legal instruction bytes are mapped to instruction
structures, which are then translated to uops.  Also, our approach is immune to
even major changes in the microcode ROM representation format, and
consequently, we are not affected by changes in the microcode ROM
assembler. This is facilitated by being able to simulate the relevant hardware
blocks by using our formal models of the SystemVerilog designs. 
Also, our uop specifications are validated not only when we prove
correspondence with EXEC, but also when we check that the
micro-programs using these uops correspond to the x86 instruction
specifications.


All of our work is done using the ACL2 theorem proving system (with
SAT used for bit-blasting), and we contribute to hardware verification
libraries that are available publicly.  Being able to analyze
commercial x86 designs using tools that are available freely
reinforces our view that formal verification of hardware in the
industry is possible without needing to purchase expensive
license-only tools.  That being said, our approach does not preclude
the use of such commercial tools; for instance, one can use 
off-the-shelf Boolean reasoning engines as backends for {\tt GL}.



The presented effort is a work in progress.  While we are close to
achieving complete verification of execution units and the decoder, we
have done single-instruction correctness proofs of only a small subset
of the 3400 x86 instruction variants in this described framework.  We
have a lot of plans for future work, both for short and long term.  As
far as verifying instruction implementations is concerned, we are
working on adding automation for tasks like proof coverage analysis.
We are also planning to extend our verification scope by focusing on
other parts of the processor, like the memory system and scheduler.

\balance

\bibliography{top}

\begin{thebibliography}{10}

\bibitem{codewalker-books}
{ACL2 Books: Codewalker}.
\newblock {Online}, Accessed: December 2019.
\newblock
  \\\url{https://github.com/acl2/acl2/tree/master/books/projects/codewalker}.

\bibitem{acl2:home}
{ACL2 Home Page}.
\newblock {Online}, Accessed: December 2019.
\newblock \\\url{http://www.cs.utexas.edu/users/moore/acl2}.

\bibitem{sv-acl2-doc}
{Documentation of {\tt SV}: A Hardware Verification Library}.
\newblock {Online}, Accessed: December 2019.
\newblock
  \\\url{http://www.cs.utexas.edu/users/moore/acl2/manuals/current/manual/?topic=ACL2____SV}.

\bibitem{vl-acl2-doc}
{Documentation of {\tt VL} Verilog Toolkit}.
\newblock {Online}, Accessed: December 2019.
\newblock
  \\\url{http://www.cs.utexas.edu/users/moore/acl2/manuals/current/manual/?topic=ACL2____VL}.

\bibitem{sv-github}
{{\tt SV}: A Hardware Verification Library}.
\newblock {Online}, Accessed: December 2019.
\newblock \\\url{https://github.com/acl2/acl2/tree/master/books/centaur/sv}.

\bibitem{svtv-doc}
{{\tt SVTV}: A Structure for Simulation Pattern of a Hardware Design}.
\newblock {Online}, Accessed: December 2019.
\newblock
  \\\url{http://www.cs.utexas.edu/users/moore/acl2/manuals/current/manual/?topic=ACL2____DEFSVTV}.

\bibitem{vl-github}
{{\tt VL} Verilog Toolkit}.
\newblock {Online}, Accessed: December 2019.
\newblock \\\url{https://github.com/acl2/acl2/tree/master/books/centaur/vl}.

\bibitem{x86isa-github}
{{\tt x86isa} Library in the ACL2 Community Books}.
\newblock {Online}, Accessed: December 2019.
\newblock
  \\\url{https://github.com/acl2/acl2/tree/master/books/projects/x86isa}.

\bibitem{hardin-centaur}
W.~{A. Hunt~Jr.}, S.~Swords, J.~Davis, and A.~Slobodova.
\newblock {Use of Formal Verification at Centaur Technology}.
\newblock In D.~Hardin, editor, {\em Design and Verification of Microprocessor
  Systems for High-Assurance Applications}, pages 65--88. Springer, 2010.

\bibitem{aagaard2003hazards}
M.~D. Aagaard.
\newblock A hazards-based correctness statement for pipelined circuits.
\newblock In {\em Advanced Research Working Conference on Correct Hardware
  Design and Verification Methods}, pages 66--80. Springer, 2003.

\bibitem{aagaard2000methodology}
M.~D. Aagaard, R.~B. Jones, T.~F. Melham, J.~W. O'Leary, and C.-J.~H. Seger.
\newblock A methodology for large-scale hardware verification.
\newblock In {\em International Conference on Formal Methods in Computer-Aided
  Design}, pages 300--319. Springer, 2000.

\bibitem{Aagaard:1998:CTP:277044.277189}
M.~D. Aagaard, R.~B. Jones, and C.-J.~H. Seger.
\newblock Combining theorem proving and trajectory evaluation in an industrial
  environment.
\newblock In {\em Proceedings of the 35th Annual Design Automation Conference},
  DAC '98, pages 538--541, New York, NY, USA, 1998. ACM.

\bibitem{aagaard1999formal}
M.~D. Aagaard, R.~B. Jones, and C.-J.~H. Seger.
\newblock Formal verification using parametric representations of boolean
  constraints.
\newblock In {\em Proceedings 1999 Design Automation Conference (Cat. No.
  99CH36361)}, pages 402--407. IEEE, 1999.

\bibitem{memocode11}
{Anna Slobodova, Jared Davis, Sol Swords, and Warren A. {Hunt~Jr.}}
\newblock {A flexible formal verification framework for industrial scale
  validation}.
\newblock In {\em {Proceedings of the $9$th IEEE/ACM International Conference
  on Formal Methods and Models for Codesign (MEMOCODE)}}, pages 89--97,
  {Cambridge, UK}, July 2011. IEEE/ACM.

\bibitem{arons2005formal}
T.~Arons, E.~Elster, L.~Fix, S.~Mador-Haim, M.~Mishaeli, J.~Shalev,
  E.~Singerman, A.~Tiemeyer, M.~Y. Vardi, and L.~D. Zuck.
\newblock Formal verification of backward compatibility of microcode.
\newblock In {\em International Conference on Computer Aided Verification},
  pages 185--198. Springer, 2005.

\bibitem{32-bit-x86isa}
A.~Coglio and S.~Goel.
\newblock Adding 32-bit mode to the acl2 model of the x86 isa.
\newblock In {\em {\rm Proceedings of the 15th International Workshop on the}
  ACL2 Theorem Prover and Its Applications, ACL2 2018, {\rm Austin, Texas, USA,
  November 5-6, 2018}}, volume 280 of {\em EPTCS}, pages 77--94, 2018.

\bibitem{centaur-itp14}
J.~Davis, A.~Slobodova, and S.~Swords.
\newblock Microcode verification -- another piece of the microprocessor
  verification puzzle.
\newblock In {\em ITP '14: Proceedings of Interactive Theorem Proving}, pages
  1--16. Springer, LNCS 8558, 2014.

\bibitem{Franzen:2010:ASS:1998496.1998520}
A.~Franz{\'e}n, A.~Cimatti, A.~Nadel, R.~Sebastiani, and J.~Shalev.
\newblock Applying smt in symbolic execution of microcode.
\newblock In {\em Proceedings of the 2010 Conference on Formal Methods in
  Computer-Aided Design}, FMCAD '10, pages 121--128, Austin, TX, 2010. FMCAD
  Inc.

\bibitem{HS09}
W.~A. {Hunt~Jr.} and S.~Swords.
\newblock {Centaur Technology} media unit verification.
\newblock In {\em Proceedings of the 21st International Conference on Computer
  Aided Verification (CAV)}, pages 353--367, 2009.

\bibitem{intel-x86-doc}
{Intel Corporation}.
\newblock {Intel$\textsuperscript{\textregistered}$ 64 and IA-32 Architectures
  Software Developer's Manual Combined Volumes: 1, 2A, 2B, 2C, 2D, 3A, 3B, 3C,
  3D, and 4}.
\newblock {Online}, May, 2019.
\newblock {Order Number: 325462-070US.
  \\\url{https://software.intel.com/en-us/articles/intel-sdm}.}

\bibitem{intel-futures-manual}
{Intel Corporation}.
\newblock {Intel$\textsuperscript{\textregistered}$ Architecture Instruction
  Set Extensions Programming Reference}, May, 2019.
\newblock {Order Number: 319433-037.
  \\\url{https://software.intel.com/en-us/articles/intel-sdm}.}

\bibitem{jones2002symbolic}
R.~B. Jones.
\newblock {\em Symbolic simulation methods for industrial formal verification}.
\newblock Springer Science \& Business Media, 2002.

\bibitem{kaivola2005formal}
R.~Kaivola.
\newblock Formal verification of pentium$\textsuperscript{\textregistered}$ 4
  components with symbolic simulation and inductive invariants.
\newblock In {\em International Conference on Computer Aided Verification},
  pages 170--184. Springer, 2005.

\bibitem{intel-i7-fv}
R.~Kaivola, R.~Ghughal, N.~Narasimhan, A.~Telfer, J.~Whittemore, S.~Pandav,
  A.~Slobodova, C.~Taylor, V.~Frolov, E.~Reeber, and A.~Naik.
\newblock Replacing testing with formal verification in
  intel$\textsuperscript{\textregistered}$ core$\textsuperscript{TM}$ i7
  processor execution engine validation.
\newblock In A.~Bouajjani and O.~Maler, editors, {\em Computer Aided
  Verification}, pages 414--429, Berlin, Heidelberg, 2009. Springer Berlin
  Heidelberg.

\bibitem{myreen-diff-arch}
M.~O. Myreen, M.~Gordon, and K.~Slind.
\newblock {Machine-Code Verification for Multiple Architectures - An
  Application of Decompilation into Logic}.
\newblock In {\em Formal Methods in Computer-Aided Design, 2008. FMCAD '08},
  pages 1--8, Nov 2008.

\bibitem{conf/cav/Reid16}
A.~Reid, R.~Chen, A.~Deligiannis, D.~Gilday, D.~Hoyes, W.~Keen, A.~Pathirane,
  E.~Shepherd, P.~Vrabel, and A.~Zaidi.
\newblock {E}nd-to-{E}nd {V}erification of {A}rm {P}rocessors with
  {I}sa-{F}ormal.
\newblock In S.~Chaudhuri and A.~Farzan, editors, {\em Proceedings of the 2016
  International Conference on Computer Aided Verification (CAV'16)}, volume
  9780 of {\em Lecture Notes in Computer Science}, pages 42--58. Springer
  Verlag, July 2016.

\bibitem{boyer1996mechanized}
{Robert S. Boyer and J S. Moore}.
\newblock {Mechanized Formal Reasoning About Programs And Computing Machines}.
\newblock {\em Automated Reasoning and Its Applications: Essays in Honor of
  Larry Wos}, pages 147--176, 1996.

\bibitem{russinoff2019formal}
D.~M. Russinoff.
\newblock {\em {Formal Verification of Floating-Point Hardware Design: A
  Mathematical Approach}}.
\newblock Springer, 2019.

\bibitem{raymoore}
{Sandip Ray and J S. Moore}.
\newblock {Proof Styles in Operational Semantics}.
\newblock In A.~J. Hu and A.~K. Martin, editors, {\em {Proceedings of the $5$th
  International Conference on Formal Methods in Computer-Aided Design (FMCAD
  2004)}}, volume 3312 of {\em {LNCS}}, pages 67--81, {Austin, TX}, Nov. 2004.
  {Springer-Verlag}.

\bibitem{ray-mechanical}
{Sandip Ray, Warren A. Hunt, Jr., John Matthews, and J S. Moore}.
\newblock {A Mechanical Analysis of Program Verification Strategies}.
\newblock {\em {Journal of Automated Reasoning}}, 40(4):245--269, {May} 2008.

\bibitem{goel-dissertation}
{Shilpi Goel}.
\newblock {\em {Formal Verification of Application and System Programs Based on
  a Validated x86 ISA Model}}.
\newblock PhD thesis, {Department of Computer Science, The University of Texas
  at Austin}, 2016.

\bibitem{spISA2019}
{Shilpi Goel and Rob Sumners}.
\newblock {Using {\tt x86isa} for Microcode Verification}.
\newblock In {\em SpISA 2019: Workshop on Instruction Set Architecture
  Specification}, 2019.

\bibitem{sat15}
A.~Slobodova.
\newblock Pragmatic approach to formal verification.
\newblock In {\em SAT '15: Proceedings of Theory And Applications Of
  Satisfiability Testing}, pages IX--XI. Springer, LNCS 9340, 2015.

\bibitem{stewart2014processor}
D.~Stewart, D.~Gilday, D.~Nevill, and T.~Roberts.
\newblock Processor memory system verification using dogrel: a language for
  specifying end-to-end properties.
\newblock In {\em International Workshop on Design and Implementation of Formal
  Tools and Systems (DIFTS)}, 2014.

\bibitem{17-swords-term-level}
S.~Swords.
\newblock Term-level reasoning in support of bit-blasting.
\newblock In A.~Slobodova and W.~A. Hunt, Jr., editors, {\em {\rm Proceedings
  14th International Workshop on the} ACL2 Theorem Prover and its Applications,
  {\rm Austin, Texas, USA, May 22-23, 2017}}, volume 249 of {\em Electronic
  Proceedings in Theoretical Computer Science}, pages 95--111. Open Publishing
  Association, 2017.

\bibitem{11-swords-bit-blasting}
S.~Swords and J.~Davis.
\newblock Bit-blasting acl2 theorems.
\newblock In D.~Hardin and J.~Schmaltz, editors, {\em {\rm Proceedings 10th
  International Workshop} on the ACL2 Theorem Prover and its Applications, {\rm
  Austin, Texas, USA, November 3-4, 2011}}, volume~70 of {\em Electronic
  Proceedings in Theoretical Computer Science}, pages 84--102. Open Publishing
  Association, 2011.

\bibitem{10-swords-dissertation}
S.~O. Swords.
\newblock {\em A Verified Framework for Symbolic Execution in the {ACL2}
  Theorem Prover}.
\newblock PhD thesis, University of Texas at Austin, December 2010.
\newblock \url{http://hdl.handle.net/2152/ETD-UT-2010-12-2210}.

\bibitem{acl2-industrial-2015}
{Warren A. Hunt, Jr., Matt Kaufmann, J S. Moore, and Anna Slobodova}.
\newblock Industrial hardware and software verification with {ACL2}.
\newblock In {\em Verified Trustworthy Software Systems}, volume 375. The Royal
  Society, 2017.
\newblock (Article Number 20150399).

\end{thebibliography}

\end{document}